# Alteration of the Brain's Microbiome and Neuroinflammation Associated with Ventricular Catheters

*Zihan Zhu, Dipankar Biswas, Michael Meggyesy, Di Cao, Gwendolyn Williams, Richard Um, Farzad Maroufi, Ryan P. Lee, Jun Hua, Liangliang Zhang, Jeffrey Capadona, Horst V. Recum, Mark G. Luciano*

## Abstract

**Background and Objectives:**

Proximal catheter obstruction is the leading cause of ventriculoperitoneal shunt failure, yet the biological triggers of peri-catheter inflammation and tissue ingrowth remain poorly defined. Emerging evidence of bacterial ribosomal RNA in human brain tissue suggests that low-biomass microbial exposure may influence the inflammatory microenvironment surrounding intracranial implants. This study examined whether microbial signal is detectable in unaltered brain tissue and whether catheter implantation produces material-dependent microbial shifts relevant to shunt dysfunction.

**Methods:**

Twenty-nine female C57BL/6 mice were assigned to unaltered control (UC), trauma control (TC), plain silicone catheter (PSC), or antibiotic-impregnated catheter (AIC) groups. Brain and cecum tissues were harvested at postoperative days 7 and 28 for 16S rRNA sequencing. Microbial composition and predicted functional pathways were analyzed. A separate cohort underwent longitudinal MRI to assess edema, glial scar formation, and macrophage-associated susceptibility signal.

**Results:**

Low-level microbial signal was detected in unaltered brain tissue. Catheter implantation induced material-dependent shifts in brain-associated microbial composition. PSC was associated with enrichment of pro-inflammatory taxa, whereas AIC favored immune-regulatory taxa. Predicted short-chain fatty acid biosynthesis was highest in AIC and lowest in PSC, while predicted lipopolysaccharide biosynthesis trended higher in PSC. MRI showed similar edema resolution but higher macrophage-associated susceptibility signal in PSC animals.

**Conclusion:**

Intracranial catheter implantation produces material-dependent shifts in low-biomass brain-associated microbial signal that parallel differential neuroimmune activation. These findings suggest catheter material may shape a biologically relevant peri-catheter niche with implications for chronic gliosis and proximal shunt obstruction.

# Introduction

Ventriculoperitoneal shunting remains the primary treatment for hydrocephalus, yet shunt failure is frequent, with proximal catheter obstruction representing the most common cause.[1,2] Although tissue ingrowth and cellular accumulation at catheter inlet pores are well-described, the biological mechanism that initiate this process remain poorly understood. In particular, the upstream inflammatory triggers that lead to catheter occlusion have not yet be clearly investigated.

Recent reports detecting bacterial ribosomal RNA in human brain tissue have renewed interest in whether the central nervous system may harbor a low-abundance microbiome.[3] Such findings have been described in immunodeficient patients and in neurodegenerative diseases,[4] raising questions about the origin and relevance of these microbial signals. Possible explanations include true microbial residency within the parenchyma, translocation through the vasculature, introduction during surgery, altered permeability of the blood-brain barrier, or contamination during sample processing or sequencing.[5–8] Clarifying the presence or absence of a native intracerebral microbiome is essential, as any microbial exposure during catheter implantation, or disruption of a preexisting microbial milieu, could shape the inflammatory environment surrounding a shunt.

# Methods

## Animals and Surgical Procedure

This study was approved by the Animal Care and Use Committee of Johns Hopkins University. Twenty-nine female C57BL/6 mice (strain #000664; 5 weeks old) were used. Animals were divided into two non-overlapping cohorts: an rRNA analysis cohort (n=24) and an MRI cohort (n=5). In the rRNA analysis cohort, mice were stratified into four equal groups (n=6 per group): 1) UC, 2) TC (no implantation), 3) Plain Silicone Catheter (PSC; Medtronic, Minneapolis, MN, USA) implanted in brain and peritoneal cavity, and 4) Antibiotic-Impregnated Catheters (AIC; Bactiseal, rifampin & clindamycin, Integra Lifesciences, Princeton, NJ, USA) implanted in brain and peritoneal cavity. In the MRI cohort, two mice were implanted with PSC fragments in the brain, two mice were implanted with AIC fragment in the brain, and one unaltered control mouse did not undergo implantation. No intraperitoneal procedures were performed for this cohort. Animals were not pre-treated with probiotics or antibiotics before procedures. Mice were housed separated by study group as defined below for the entirety of the study duration.

Sterile PSC and AIC fragments were hand-cut roughly as a rectangular parallelopiped with dimensions 1.00mm x 1.00mm x 0.65mm. All fragments were sterilized using ethylene oxide gas in a conventional hospital sterilizer. Microsurgical sets were sterilized between each animal, and separate sterile sets were reserved for each implantation group to prevent cross-contamination.

All mice were sedated by an intraperitoneal injection of a ketamine/xylazine/ethanol mixture. Following sedation, mice were place in the prone position and a midline cranial scalp incision was made. A burr hole was placed in the right parietal bone and a PISC or AIC fragment was implanted intra-parenchymally just below the cortex, avoiding ventricular contact. For the positive trauma control group the PSC fragment was immediately removed following placement.

For the rRNA analysis cohort only, a second procedure was performed in the supine position. A small midline abdominal incision was made to access the peritoneal cavity. The respective catheter fragment for each group was implanted (or mock-implanted for the TC) in the peritoneum near the cecum. All incisions were closed with sutures and mice were monitored post-operatively.

**16S rRNA Sequencing and Analysis**

Mice were sacrificed at two postoperative time points: acute (day 7; n = 3 per group) and chronic (day 28; n = 3 per group). Brain and cecum tissues were harvested, flash-frozen, and stored at −80°C until processing. DNA was extracted from brain and cecum tissue and used for 16S rRNA gene sequencing. Libraries were prepared using the Earth Microbiome Project protocol and sequenced on an Illumina MiSeq platform (paired-end 150 bp).[9] Sequence data were processed using QIIME 2 (v2024.5) with host-read removal and denoising prior to taxonomic assignment.

Microbial community structure was evaluated using alpha diversity (Shannon index, Chao1 richness) and beta diversity (weighted UniFrac). Group-wise comparisons were performed using Wilcoxon rank-sum tests (alpha diversity) and PERMANOVA (beta diversity).

Differential abundance was assessed using Linear discriminant analysis Effect Size (LEfSe), with an emphasis on effect size using linear discriminant analysis (LDA) score rather than statistical significance due to limited sample size. Functional potential was inferred using PICRUSt2, focusing on short-chain fatty acid (SCFA) and lipopolysaccharide (LPS) biosynthesis pathways. Effect sizes were quantified using Cliff's delta.

Additional experimental, sequencing, bioinformatics, and statistical details are provided in the Supplemental Methods.

**Sequence Processing and Taxonomic Classification**

Host-derived reads were removed prior to microbial analysis using Kraken2 (v2.1.1.2).[10] High-quality microbial reads were denoised and clustered into operational taxonomic units (OTUs) at 99% identity against the SILVA reference database.[11] Taxonomic classification was performed using a Naive Bayes classifier.

Further details regarding quality control, read filtering, clustering parameters, and reference databases are provided in the Supplemental Methods.

**Statistical Analysis**

All statistical analyses were conducted in R (v4.4.1). Alpha diversity was assessed using the Shannon diversity index and Chao1 richness estimator. The Shannon index was computed using the diversity() function and Chao1 values were obtained via the estimateR() function from the vegan package (v2.6-4).[12] For group-wise comparisons, Wilcoxon rank-sum tests were used to evaluate differences in alpha diversity between brain and cecum samples, as well as between acute and chronic samples. Beta diversity was assessed using weighted UniFrac distances[13,14] and tested using PERMANOVA. Differential abundance was evaluated using LEfSe.[15] Functional profiles of the microbial communities were predicted from 16S rRNA gene sequences using PICRUSt2.[16] Within both the brain and cecum subsets, we compared the predicted abundances of these functional pathways among unaltered control, trauma control, PSC implantation, and AIC implantation groups using boxplots. Effect size is calculated using Cliff's Delta and a permutation p-value is provided.[17] Expanded statistical methods, including software versions and package details, are provided in the Supplemental Methods.

**In Vivo MR Imaging and Data Analysis**

In vivo MRI was performed on five mice: negative control (n = 1), PSC-implanted (n = 2), and AIC-implanted (n = 2), at 1, 4, 8, and 16 weeks post-implantation. Anatomical T2-weighted, FLAIR, and susceptibility-weighted imaging (SWI) sequences were acquired. Ferumoxytol was administered prior to SWI to enhance macrophage-associated iron contrast. Four regions of interest (ROIs) were defined: implant, edema, glial scar, and contralateral control tissue. ROI volumes and R2* values were quantified to estimate tissue response and macrophage activity. Given the small sample size, MRI-derived endpoints were interpreted descriptively rather than through formal hypothesis testing. Further MRI acquisition details and image processing and ROI delineation criteria are provided in the Supplemental Methods.

**Results**

**16S rRNA Analysis**

We first examined microbial presence and overlap across anatomical sites in unaltered control (UC) mice using operational taxonomic unit (OTU) profiles. To increase statistical power, data from day 7 and day 28 were pooled. A total of 44 OTUs were shared between the brain and cecum, while 55 and 191 OTUs were unique to the brain and cecum, respectively (Figure 6a). A relative abundance heatmap of individual UC brain samples demonstrated low-abundance but consistent detection of genera including Cutibacterium, Flavobacterium, Pseudomonas, Muribaculaceae, and Herbaspirillum (Figure 6b).

Next, we assessed microbial overlap across all mice irrespective of treatment group or timepoint. Across the full dataset, 84 OTUs were shared between brain and cecum samples, while 95 and 151 OTUs were unique to the brain and cecum, respectively (Figure 6c). These findings indicate substantial site-specific microbial differentiation, with the peritoneal compartment exhibiting greater taxonomic richness.

To further assess how microbial presence varied across both anatomical location and inflammatory stage, samples were stratified into Brain–Acute (BA), Brain–Chronic (BC), Cecum–Acute (CA), and Cecum–Chronic (CC) groups. The BC group exhibited the highest number of unique OTUs (n = 16), followed by CC (12), CA (9), and BA (4) (Figure 6d). A core set of 23 OTUs was present across all four subgroups. In contrast, 75 OTUs were shared only between BA and BC samples, while 130 OTUs were shared only between CA and CC samples, reinforcing strong site-dependent microbial patterning.

Chao1 richness indices confirmed that the cecum harbored significantly more taxa than the brain ($p < 0.0001$), consistent with the greater OTU richness observed in Venn analyses. In contrast, Shannon diversity did not differ significantly between sites ($p = 0.8$), suggesting the brain, compared to cecum, has fewer separate species, but a more even bacterial abundance distributed among each species. Overall, the brain population is considered less stable (Figure 7a). Principal coordinates analysis (PCoA) based on weighted UniFrac distances demonstrated clear separation between brain and peritoneal microbiota (PERMANOVA p = 0.001; Figure 7b), indicating distinct phylogenetic structures by anatomical site.

To assess whether inflammation stage also contributed to microbial variation, we performed additional alpha and beta diversity comparisons between acute and chronic samples within each body site. No significant differences were observed, suggesting that microbial diversity was shaped primarily by anatomical context rather than temporal progression. Based on these site-dominant patterns, brain and cecum samples were analyzed separately in all subsequent analyses.

**Brain Microbiome After Intervention**

Two microbes, Gammaproteobacteria and Saccharimonadales, were enriched in the TC group as seen in Figure 8a. The former comprises several clinically relevant opportunistic pathogens, while the later, a member of the Candidate Phyla Radiation represents a group of ultrasmall, co-dependent bacteria often emerging under dysbiotic conditions.

Subgroup-specific LEfSe analyses were performed to compare each treatment group (TC, PSC, and AIC) separately against the negative control. Taxa enriched across multiple comparisons were assigned to the group in which enrichment was strongest. The resulting discriminative taxa are shown in Figure 8a, ranked by linear discriminant analysis (LDA) score. The PSC group was enriched in Desulfovibrionaceae, Muribaculaceae, and Clostridia UCG-014 many of which have been implicated in pro-inflammatory processes, intestinal barrier dysfunction, or the production of lipopolysaccharide (LPS).[18–21] In contrast, the AIC group showed enrichment of Akkermansiaceae (notably Akkermansia), Parabacteroides, and unclassified Clostridiales; all of which have been linked to the production of short-chain fatty acids (SCFAs), epithelial barrier maintenance, and reduced neuroinflammation in both gut and brain studies.[22–25] The TC group was dominated by Providencia, Saccharimonadales, and Gammaproteobacteria, which are often seen in response to external stress, surgical trauma, or infection-related dysbiosis.[26–29]

Several taxa were consistently enriched across more than one implant group. Providencia (Morganellaceae) was enriched across all three groups. Enhydrobacter and members of the Burkholderiaceae family were enriched in the TC and PSC groups but not in the AIC group. Conversely, Gammaproteobacteria and Saccharimonadales were enriched in the TC and AIC groups but not in the PSC group.

**Predicted Functional Profiles in the Brain**

Given these taxonomic contrasts, predicted microbial functional profiles were assessed using PICRUSt2, focusing on short-chain fatty acid (SCFA) and lipopolysaccharide (LPS) biosynthesis pathways (Figures 8b–c). SCFA pathway abundance was highest in the AIC group, intermediate in the TC group, and lowest in the PSC group. Compared with the negative control, AIC exhibited a large positive effect size (Cliff's $\Delta = 0.76$, p = 0.0524), whereas PSC showed minimal change ($\Delta = 0.133$, p = 0.7898). The AIC vs. PSC comparison yielded $\Delta = -0.733$ (p = 0.0513), indicating a strong trend toward increased SCFA biosynthetic potential in AIC-associated microbiota.

Predicted LPS biosynthesis pathway abundance showed smaller, nonsignificant differences. Directionally, AIC had lower LPS potential than both PSC (Δ = −0.44) and UC (Δ = −0.44), while PSC exhibited the highest LPS levels (Δ = 0.4 vs. UC) (Figure 8c).

**Cecum Microbiome**

In the TC group we observed the abundance of Streptococcus—a genus frequently associated with epithelial barrier disruption and post-surgical dysbiosis—and Tuzzerella, an emerging Lachnospiraceae lineage that has been observed to expand following antibiotic exposure or microbial perturbation.[30,31] LEfSe analysis identified several discriminative taxa across the intervention groups (Figure 9a). The PSC group was enriched in Oscillospirales, RF39, and the Clostridia vadinBB60 group, which have been associated with altered fermentation dynamics and microbiome restructuring in inflammatory or antibiotic-exposed contexts.[32,33] The AIC group showed relative enrichment of Lachnospiraceae UCG-001 and Deferribacterota, which are known to colonize low-oxygen mucosal niches and may play a role in ecological stabilization following microbial perturbation.[34,35] Several taxa were consistently enriched across multiple groups. Deferribacterota, Mucispirillum, and Ruminococcaceae were frequently detected. Streptococcus and Tuzzerella were enriched in both AIC and TC groups. The Clostridia vadinBB60 group appeared across all comparisons.

Predicted SCFA pathway abundance was broadly similar across groups, with no statistically significant differences (Figure 9b). AIC exhibited slightly higher SCFA potential compared with PSC and negative control, although effect sizes were small (Δ = 0.267, p = 0.5376 vs. negative control; Δ = 0.8, p = 0.0646 vs. PSC). Predicted LPS pathway abundance also showed minimal differentiation (Figure 9c). AIC showed higher predicted LPS pathway abundance than both negative control (Δ = 0.267, p = 0.5393) and PSC (Δ = –0.067, p = 0.9186), though again, statistical support was limited.

**Temporal Dynamics: Acute Versus Chronic**

To assess time-dependent microbial shifts, we analyzed $\log_2$ fold changes in genus-level abundance between day 7 and day 28 across all treatment groups and body sites (Figure 10). In brain samples, the AIC group exhibited early enrichment of Akkermansia, Bacteroides, and Turicibacter. In contrast, PSC and TC groups showed early increases in Providencia, Saccharimonadales, and Tannerellaceae.

Providencia exhibited the strongest expansion in the AIC group. In contrast to the brain, peritoneal microbiota remained comparatively stable across all groups, with modest fold changes and preservation of genera such as Lachnospiraceae NK4A136 group and Lactobacillus.

**MRI Results**

MRI analysis confirmed reliable identification and quantification of implanted catheter fragments. Catheter volumes were consistently slightly below 1 mm³ across all implanted mice and timepoints (Figure 4a). R2*, the relaxation rate that quantifies iron overload within the implant region, showed no significant changes over time (Figure 5a).

Representative T2-weighted and FLAIR images at 1 and 16 weeks post-implantation are shown in Figure 3. Edema volumes peaked at week 1, declined sharply by week 4, and remained negligible at weeks 8 and 16 (Figure 4). Glial scar volumes were minimal at week 1, peaked at week 4, and persisted through week 16. No significant volumetric differences were observed between PSC- and AIC-implanted mice.

Macrophage activity, assessed via R2* changes in susceptibility-weighted imaging (Figure 5), increased in glial scar regions from week 1 to week 8 and remained elevated through week 16. R2* values in glial scar regions were consistently higher in PSC than in AIC mice at weeks 4, 8, and 16. Contralateral control regions showed stable R2* values across all groups and timepoints. Glial scar R2* values were higher than contralateral tissue at all timepoints in all implanted mice.

**Discussion**

In this pilot study, we evaluated whether microbial signal is detectable in unaltered brain tissue and whether intracranial catheter implantation is associated with measurable shifts in microbial community structure. Using 16S rRNA sequencing across multiple anatomical sites, time points, and implant conditions—together with longitudinal MRI—we identified reproducible, directionally consistent microbial signatures in brain tissue and showed that these signatures differed by catheter material and evolved over time.

Across all experimental groups, including unaltered controls, we detected low-level microbial signal in brain samples. This finding aligns with prior reports describing microbial DNA or RNA in tissues traditionally considered sterile[36–39] and supports the possibility that the brain harbors either a transient microbial trace population or persistent microbial fragments derived from circulation, meningeal interfaces, or glymphatic trafficking. Importantly, however, the microbial profiles observed here were not random: they exhibited consistent taxonomic structure, reproducibility across animals, and systematic modulation by implant material and time point. These features argue against stochastic contamination as a sole explanation, although low-biomass sequencing artifacts remain an unavoidable methodological concern.

Against this baseline, catheter implantation induced substantial and material-dependent shifts in brain-associated microbial composition. PSC and AIC, are the ones most commonly used for shunt implantations around the world. AICs have higher associated material costs, and studies have shown a tendency towards lower infection rates.[40] This trend however could not be verified in different populations[41] and remains a topic of research. Some institutions reserve AIC for complicated cases only. AIC catheters release their antibiotics within the first 28 days but provide a more prolonged protection against certain bacterial strains up to 56 days.[42] PSC implantation was associated with enrichment of taxa such as Desulfovibrionaceae, Muribaculaceae, and Clostridia UCG-014, several of which have been linked in other systems to endotoxin production, intestinal barrier disruption, or pro-inflammatory signaling. In contrast, AIC implantation was associated with enrichment of taxa including Akkermansia, Parabacteroides, and unclassified Clostridiales, organisms more commonly associated with short-chain fatty acid (SCFA) production, epithelial integrity, and immune-regulatory effects in gut–brain axis studies. These contrasting microbial signatures suggest that catheter material and antibiotic impregnation shape the local ecological niche in a non-neutral fashion, potentially favoring either dysbiosis-associated or immunomodulatory microbial communities.

Functional inference using PICRUSt2 reinforced this taxonomic dichotomy. Predicted SCFA biosynthetic potential was highest in the AIC group and lowest in the PSC group, whereas predicted LPS biosynthesis potential showed the opposite directional trend. Although these pathway estimates are indirect and require experimental validation, the coherence between taxonomic enrichment (SCFA-associated vs. endotoxin-associated taxa) and predicted functional directionality strengthens the biological plausibility of a material-driven immunological contrast. In this framework, AICs may foster a relatively anti-inflammatory microbial milieu, while PSCs may permit expansion of taxa more closely linked to endotoxin signaling and neuroinflammatory activation.

Temporal analyses further suggested that these effects are dynamic rather than static. In brain samples, AIC implantation was associated with early enrichment of Akkermansia, Bacteroides, and Turicibacter—genera linked to mucosal support and SCFA metabolism—during the acute phase. In contrast, PSC and trauma control groups exhibited sustained or increasing abundance of taxa such as Providencia, Saccharimonadales, and Tannerellaceae, which are frequently observed in dysbiosis-associated contexts. These findings imply that AICs may transiently stabilize the early post-implant microbiome toward a potentially protective configuration, whereas PSCs may permit early dysbiotic shifts that persist into the chronic phase.

One of the most clinically salient findings was the expansion of Providencia in the AIC group despite antibiotic impregnation. Providencia species are facultative anaerobic Gram-negative organisms known for intrinsic resistance to clindamycin and rapid acquisition of rifampicin resistance through single-point mutations in rpoB. Their marked expansion under antibiotic pressure suggests an "antibiotic escape" phenomenon in the low-biomass intracranial niche. This observation highlights a fundamental limitation of current antimicrobial catheter strategies, which are optimized primarily against Gram-positive organisms and may inadvertently select for resilient Gram-negative taxa.[43] These data argue for future biomaterial designs that either broaden antimicrobial spectrum or incorporate ecological stabilization strategies—such as promoting SCFA-producing taxa that may suppress opportunistic pathogens through niche competition.

In contrast to the brain, the peritoneal compartment exhibited higher baseline OTU richness, broader taxonomic diversity, and substantially greater temporal stability. Across taxonomic, functional, and temporal analyses, peritoneal microbial profiles showed muted responses to implant material and minimal divergence between acute and chronic phases. Predicted SCFA and LPS biosynthesis pathways differed only modestly across groups, and fold-change dynamics were constrained relative to those observed in brain tissue. These findings align with long-standing clinical experience that the peritoneal cavity is a microbiologically permissive and immunologically buffered environment for cerebrospinal fluid diversion.[44–46] In this model, the peritoneum appears to function as an ecological sink that absorbs microbial perturbation without manifesting pronounced dysbiosis, whereas the brain exhibits selective, dynamic, and material-sensitive microbial remodeling.

MRI provided complementary longitudinal context for these microbiome findings. As stated in previously published literature, post-implantation edema starts on day-1 and lasts approximately four to eight days[47] while glial scarring is expected to start forming at two-weeks post-implantation completing at four to six weeks. As expected, edema peaked early after implantation and resolved, while glial scar volume increased and persisted through 16 weeks. Although volumetric measures did not differ substantially between PSC and AIC groups, susceptibility-based imaging revealed greater macrophage-associated signal in glial scar regions of PSC-implanted mice at later time points. While the MRI cohort was small and not powered for inferential statistics, the directionality of this signal is consistent with the broader microbiome findings: PSCs were associated with enrichment of pro-inflammatory taxa and higher predicted LPS potential, whereas AICs were associated with enrichment of SCFA-linked taxa and lower predicted LPS potential. This convergence raises the possibility that implant-associated microbial shifts contribute, directly or indirectly, to chronic neuroimmune activation and glial scar remodeling.

**Limitations**

This pilot study is limited primarily by its small sample size, which restricts statistical power and generalizability. Nevertheless, these preliminary findings support further investment in larger-scale studies to more rigorously test these hypotheses. More refined taxonomic subclassification of microbial species would also improve interpretability by clarifying which organisms are involved and how they may influence immune and inflammatory responses.

Despite adherence to best practices, the possibility of sample contamination or detection of background "kitome" 16S rRNA sequences must be acknowledged. Minor compositional changes in a negative-control mouse may reflect technical variability, background temporal shifts, or incomplete evacuation of intravascular blood from cerebral vessels, allowing circulating microbes to be detected in brain tissue. This limitation could be addressed by performing paired 16S rRNA analyses on blood samples from the same animals and using these as internal controls. Similarly, small-magnitude changes were observed in cecal samples, where large effects were not anticipated, potentially due to spatial separation between the catheter fragment and analyzed tissue.

Another limitation is that the implanted model did not represent a fully functioning shunt system. Future studies should incorporate operational shunts and integrate proteomic, histological, and immunohistochemical analyses to better approximate clinical conditions and assess gliosis.

Taken together, these data support a conceptual model in which intracranial catheter materials shape a low-biomass but biologically responsive microbial niche in the brain, with downstream implications for neuroinflammation. Whether these microbial shifts are causal drivers of tissue response, passive markers of host immune state, or epiphenomena of a perturbed local ecosystem remains unresolved. Nonetheless, the coherence across taxonomic, functional, temporal, and imaging-based endpoints provides a rationale for larger studies designed to directly test microbial–immune–material interactions.

**Conclusion**

Intracranial catheter implantation was associated with measurable, material-dependent shifts in brain-associated microbial signal in a murine model. Compared with trauma controls, PSC and AIC implants produced larger changes in taxonomic profiles over time, suggesting an effect beyond surgical injury alone. Predicted functional profiles differed by implant type, with AIC-associated communities showing increased SCFA biosynthetic potential and reduced LPS biosynthesis potential relative to PSC. Longitudinal MRI demonstrated expected edema resolution and persistent glial scar formation, with higher macrophage-associated susceptibility signal in PSC animals at later time points.

Although limited by small sample size and the constraints of low-biomass microbiome profiling, these findings support further investigation of host–microbe–material interactions around intracranial implants. Larger studies incorporating higher-resolution microbial profiling and direct functional validation are warranted to determine whether material-driven microbial shifts contribute to neuroinflammation and clinically relevant outcomes such as catheter obstruction and shunt malfunction.

**Conflict of Interests:**

There is no conflict of interest

25. Atarashi K, Tanoue T, Oshima K. Treg induction by a rationally selected mixture of Clostridia strains from the human microbiota. *Nature*. 2013;500(7461):232-236. doi:10.1038/nature12331

26. Pallett SJC, Morkowska A, Woolley SD. Evolving antimicrobial resistance of extensively drug-resistant Gram-negative severe infections associated with conflict wounds in Ukraine. *Lancet Regional Health – Europe*. 2025;52. doi:10.1016/j.lanepe.2025.101274

27. Malviya M, Kale-Pradhan P, Coyle M, Giuliano C, Johnson LB. Clinical and Drug Resistance Characteristics of Providencia Infections. *Microorganisms*. 2024;12(10). doi:10.3390/microorganisms12102085

28. McLean JS, Bor B, Kerns KA. Acquisition and Adaptation of Ultra-small Parasitic Reduced Genome Bacteria to Mammalian Hosts. *Cell Rep*. 2020;32(3). doi:10.1016/j.celrep.2020.107939

29. Guyton K, Alverdy JC. The gut microbiota and gastrointestinal surgery. *Nat Rev Gastroenterol Hepatol*. 2017;14(1):43-54. doi:10.1038/nrgastro.2016.139

30. Fu K, Cheung AHK, Wong CC. Streptococcus anginosus promotes gastric inflammation, atrophy, and tumorigenesis in mice. *Cell*. 2024;187(4):882-896. doi:10.1016/j.cell.2024.01.004

31. Cao P, Yue M, Cheng Y. Naringenin prevents non-alcoholic steatohepatitis by modulating the host metabolome and intestinal microbiome in MCD diet-fed mice. *Food Sci Nutr*. 2023;11(12):7826-7840. doi:10.1002/fsn3.3700

32. Chen YJ, Ho HJ, Tseng CH. Short-chain fatty acids ameliorate imiquimod-induced skin thickening and IL-17 levels and alter gut microbiota in mice. *Sci Rep*. 2024;14(1). doi:10.1038/s41598-024-67325-x

33. Hamad I, Van Broeckhoven J, Cardilli A. Effects of Recombinant IL-13 Treatment on Gut Microbiota Composition and Functional Recovery after Hemisection Spinal Cord Injury in Mice. *Nutrients*. 2023;15(19). doi:10.3390/nu15194184

34. Qi L, Shi M, Zhu FC, Lian CA, He LS. Genomic evidence for the first symbiotic Deferribacterota, a novel gut symbiont from the deep-sea hydrothermal vent shrimp Rimicaris kairei. *Front Microbiol*. 2023;14. doi:10.3389/fmicb.2023.1179935

35. Molino S, Lerma-Aguilera A, Jimenez-Hernandez N, Rufian Henares JA, Francino MP. Evaluation of the Effects of a Short Supplementation With Tannins on the Gut Microbiota of Healthy Subjects. *Front Microbiol*. 2022;13. doi:10.3389/fmicb.2022.848611

36. Branton WG, Lu JQ, Surette MG. Brain microbiota disruption within inflammatory demyelinating lesions in multiple sclerosis. *Sci Rep*. 2016;6. doi:10.1038/srep37344

37. Emery DC, Shoemark DK, Batstone TE. 16S rRNA Next Generation Sequencing Analysis Shows Bacteria in Alzheimer's Post-Mortem Brain. *Front Aging Neurosci*. 2017;9. doi:10.3389/fnagi.2017.00195

38. Alonso R, Pisa D, Fernandez-Fernandez AM, Carrasco L. Infection of Fungi and Bacteria in Brain Tissue From Elderly Persons and Patients With Alzheimer's Disease. *Front Aging Neurosci*. 2018;10. doi:10.3389/fnagi.2018.00159
12

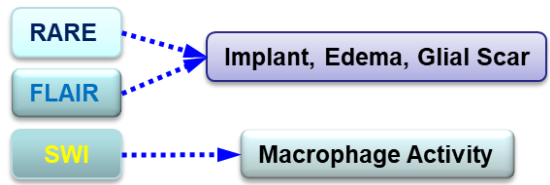

*Figure 1: Schematic representation of MR image analysis.*

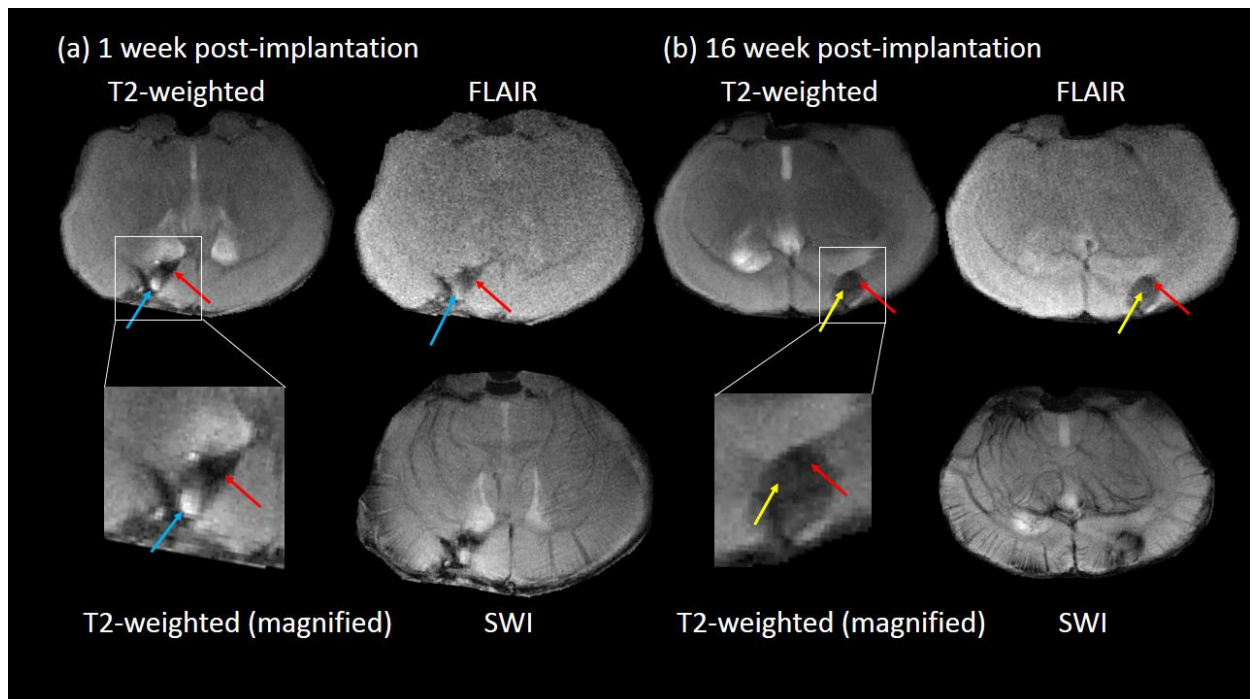

*Figure 2: Method of identification of regions of interest following intensity difference.*

*Figure 3: Representative T2-weighted, FLAIR and SWI MRI images. **(a)** MRI images from a mouse implanted with the standard catheter acquired at 1 week post-implantation. The T2-weighted image in the white box is magnified for better visualization. Red arrow: ROI; blue arrow: edema. **(b)** MRI images from a mouse implanted with the AIC catheter acquired at 16 weeks post-implantation. Red arrow: implant; yellow arrow: glial scar.*

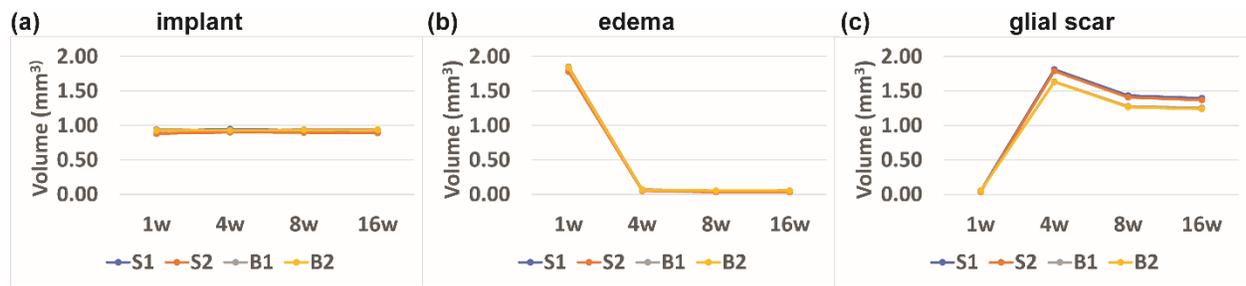

*Figure 4: Longitudinal volumetric changes in (a) the implant ROI, (b) the edema ROI, and (c) the glial scar ROI, comparing between plain silicone (S) and antibiotic-impregnated (B) catheters. The volumes of each region of interest were measured in each mouse at 1, 4, 8, and 16 weeks after the catheters were implanted. S1, S2: mice with standard catheter. B1, B2: mice with AIC catheter. "w" = week.*

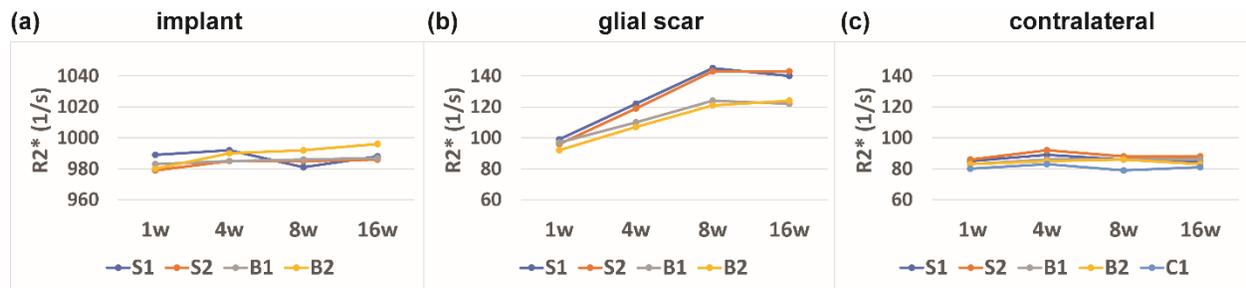

*Figure 5: Longitudinal R2\* changes—a measure of macrophage activity—in (a) the implant ROI, (b) the glial scar ROI, and (c) the contralateral tissue ROI. The R2\* values were measured in each mouse at 1, 4, 8, and 16 weeks after the catheters were implanted. S1, S2: mice with standard catheter. B1, B2: mice with AIC catheter. C1: control mouse with sham implantation.*

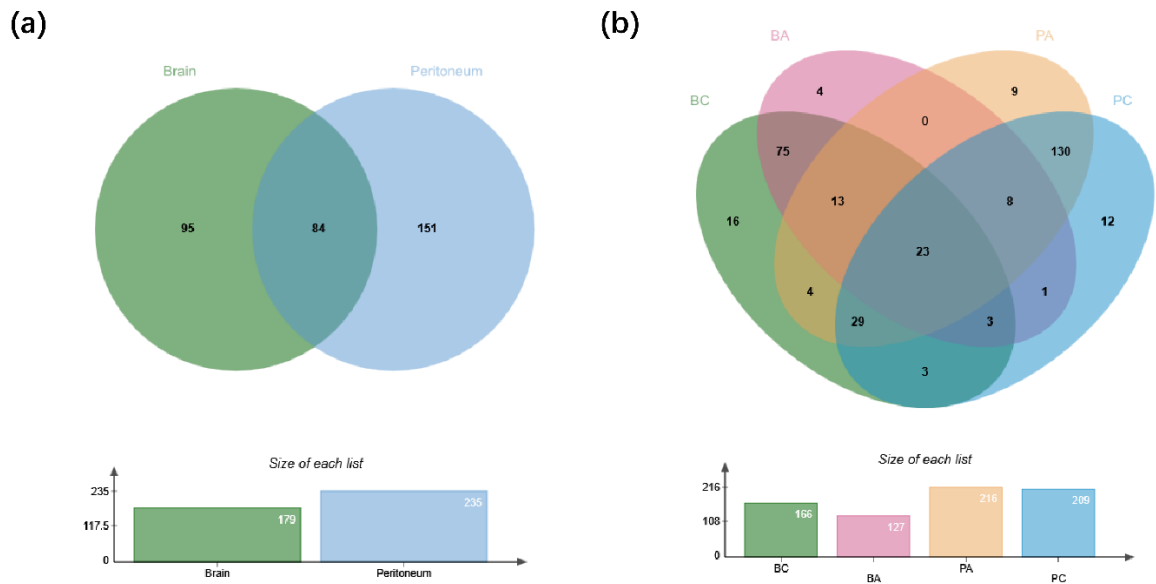

*Figure 6. **Overview of microbial community composition and OTU overlap across experimental conditions.***
*(a) Venn diagram showing shared and unique operational taxonomic units (OTUs) between brain and Cacum samples across all mice.*
*(b) Venn diagram illustrating OTU overlap across four experimental subgroups: Brain–Chronic (**BC**), Brain–Acute (**BA**), Cacum–Chronic (**PC**), and Cacum–Acute (**PA**). Each subset reflects unique or shared microbial features, highlighting both anatomical and inflammatory stage-specific microbial compositions. The adjacent bar plots indicate the total number of OTUs detected in each subgroup.*

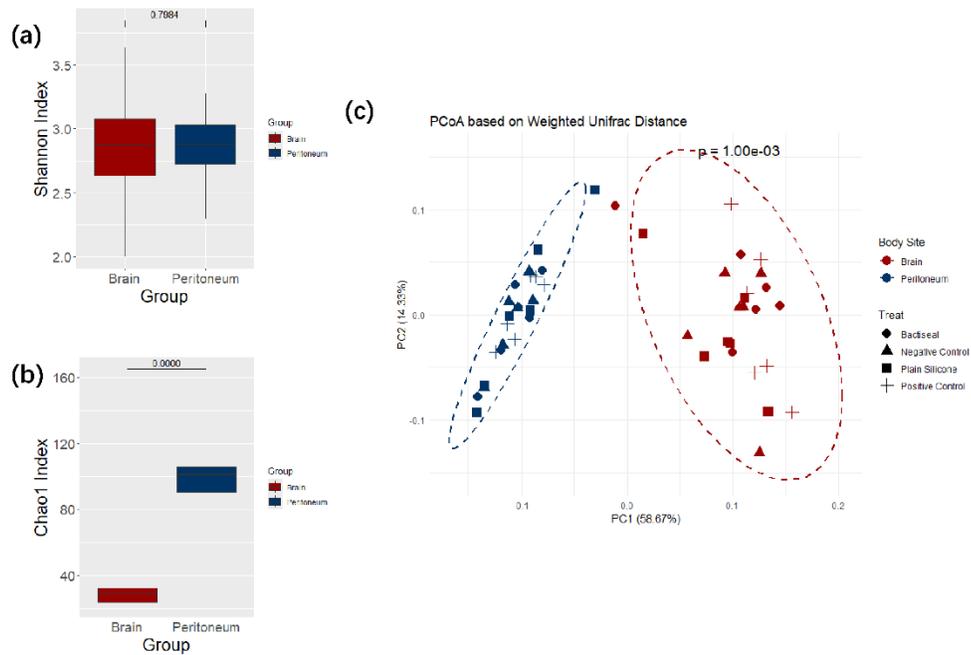

*Figure 7. Comparison of microbial diversity and community structure between brain and Cacum.*
*(a) Shannon diversity index did not differ significantly between brain and Cacum samples (p = 0.798).*
*(b) Chao1 richness index was significantly lower in brain samples compared to Cacum samples (p < 0.0001), indicating reduced species richness in the brain.*
*(c) Principal Coordinates Analysis (PCoA) based on weighted UniFrac distances revealed clear separation between brain and Cacum microbiota (PERMANOVA p = 0.001). Each point represents a mouse sample, colored by body site and shaped by treatment group. Ellipses denote 95% confidence intervals for each body site group.*

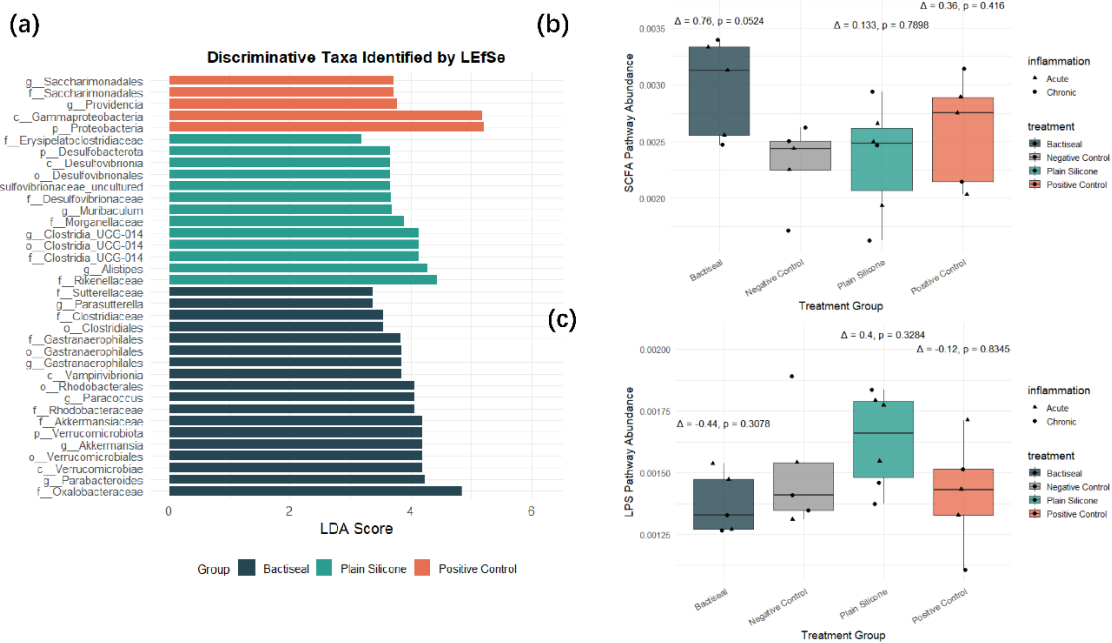

*Figure 8.*
*Discriminant microbial taxa and predicted functional pathways across implant groups. for Brain samples.*
*(a) LEfSe analysis identified microbial taxa that discriminated of AIC, Plain Silicone, and Positive Control groups against Negative Control group. Bars represent the linear discriminant analysis (LDA) scores for taxa with LDA > 2.0. Taxa are colored by group of enrichment.*
*(b) Predicted abundance of the short-chain fatty acid (SCFA) biosynthesis pathway across treatment groups based on PICRUSt2 functional inference. Each point represents a sample, shaped by inflammation status (acute = triangle; chronic = circle). Cliff's Delta (Δ) and permutation-based p-values are shown above each comparison.*
*(c) Predicted abundance of the lipopolysaccharide (LPS) biosynthesis pathway shows moderate variation across groups, but no significant differences were detected.*

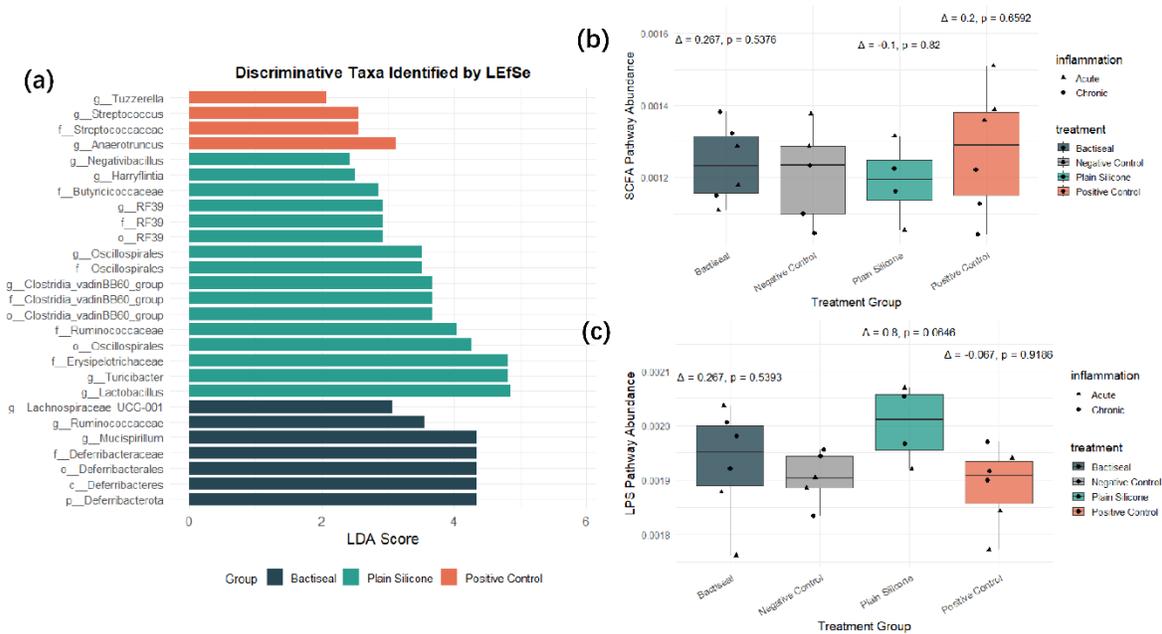

*Figure 9.*
*Discriminant microbial taxa and predicted functional pathways across implant groups for Cacum samples.*
*(a) LEfSe analysis identified microbial taxa that discriminated of AIC, Plain Silicone, and Positive Control groups against Negative Control group. Bars represent the linear discriminant analysis (LDA) scores for taxa with LDA > 2.0. Taxa are colored by group of enrichment.*
*(b) Predicted abundance of the short-chain fatty acid (SCFA) biosynthesis pathway across treatment groups based on PICRUSt2 functional inference. Each point represents a sample, shaped by inflammation status (acute = triangle; chronic = circle). Cliff's Delta (Δ) and permutation-based p-values are shown above each comparison.*
*(c) Predicted abundance of the lipopolysaccharide (LPS) biosynthesis pathway shows moderate variation across groups, but no significant differences were detected.*

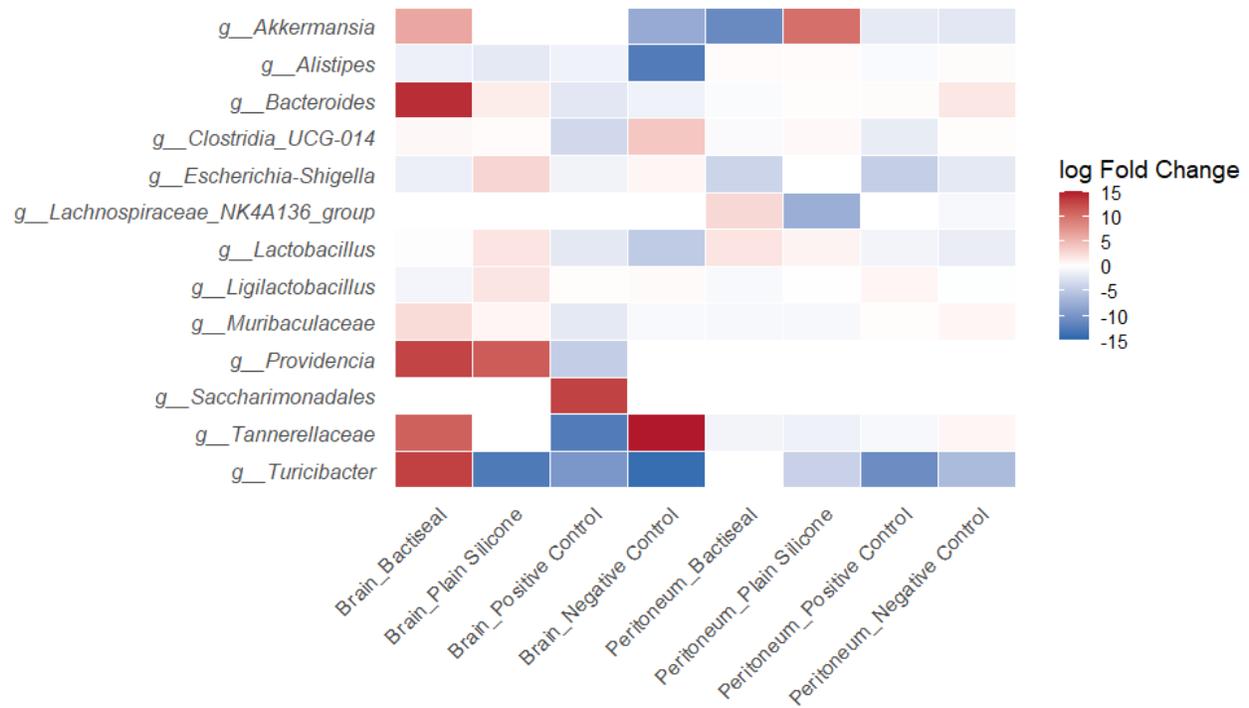

*Figure 10: Temporal dynamics of representative microbial genera across treatment groups. Shown are microbial genera with the largest changes in relative abundance between day 7 (acute inflammation) and day 28 (chronic inflammation), calculated separately within each treatment group and body site (Brain or Cacum). Genera were selected based on having the highest overall temporal fold changes across groups, excluding likely contaminants. Colors represent the log$_2$ fold change in mean relative abundance between day 28 and day 7. Positive values (red) indicate genera that expanded from day 7 to day 28, while negative values (blue) indicate genera that decreased over time. This visualization highlights both inflammation-associated microbial responses and putative recovery signatures under different implant conditions.*

**Supplemental Methods**

**Animals and Surgical Procedure**

This study was approved by the Animal Care and Use Committee of Johns Hopkins University. Twenty-nine female C57BL/6 mice (strain #000664; 5 weeks old) were used. Animals were divided into two non-overlapping cohorts: an rRNA analysis cohort (n=24) and an MRI cohort (n=5). In the rRNA analysis cohort, mice were stratified into four equal groups (n=6 per group): 1) UC, 2) TC (no implantation), 3) Plain Silicone Catheter (PSC; Medtronic, Minneapolis, MN, USA) Implanted in brain and peritoneal cavity, and 4) Antibiotic-Impregnated Catheters (AIC; Bactiseal, rifampin & clindamycin, Integra Lifesciences, Princeton, NJ, USA) Implanted in brain and peritoneal cavity. In the MRI cohort, two mice were implanted with PSC fragments in the brain, two mice were implanted with AIC fragment in the brain, and one unaltered control mouse did not undergo implantation. No intraperitoneal procedures were performed for this cohort. Animals were not pre-treated with probiotics or antibiotics before procedures. Mice were housed separated by study group as defined below for the entirety of the study duration. Sterile PSC and AIC fragments were hand-cut roughly as a rectangular parallelopiped with dimensions 1.00mm x 1.00mm x 0.65mm. All fragments were sterilized using ethylene oxide gas in a conventional hospital sterilizer. Microsurgical sets were sterilized between each animal, and separate sterile sets were reserved for each implantation group to prevent cross-contamination.

All mice were sedated by an intraperitoneal injection of a ketamine/xylazine/ethanol mixture ("Xylanest"). Following sedation, mice were place in the prone position and a midline cranial scalp incision was made. A burr hole was placed in the right parietal bone and a PISC or AIC fragment was implanted intra-parenchymally just below the cortex, avoiding ventricular contact. For the positive trauma control group the PSC fragment was immediately removed following placement.

For the RNA analysis cohort only, a second procedure was performed in the supine position. A small midline abdominal incision was made to access the peritoneal cavity. The respective catheter fragment for each group was implanted (or mock-implanted for the TC) in the peritoneum near the cecum. All incisions were closed with sutures and mice were monitored post-operatively.

**16S rRNA Sequencing and Analysis**

For the rRNA analysis cohort, mice were sacrificed for analysis at two different postoperative time points: acute (day-7; n=3/group) and chronic (day-28; n=3/group). Mice were euthanized by intracardiac injection of 0.3 ml Xylanest/Bupivacain, followed by intracardiac perfusion of 25 ml PBS over 10 minutes, then 25 ml of 30% (w/v) Sucrose solution over 10 minutes. After explantation, the brain and cecum were immediately placed in a -80°C freezer until further analysis.

Each mouse was implanted with catheter fragments unilaterally in the brain. Near implant brain tissue was used for bulk nCounter analysis. Cecum analysis method[….?] Excised brain tissue samples were directly placed in microcentrifuge tubes containing Qiazol (RNA extraction lysate). The extracted tissue was homogenized in Qiazol using 1.5 mm zirconium beads with the Bead Bug Homogenizer at high speed. RNA was immediately extracted and purified using the RNeasy Plus Universal Mini Kit following the manufacturer's protocol. RNA was stored at -80°C until further processing. RNA purity and concentrations were determined using the Qubit Fluorometer.

DNA was isolated from mouse brain tissue using QIAGEN DNA Extraction Kits. 16S-rRNA libraries were prepared and sequenced on the Illumina MiSeq platform at the Genomics core in the CWRU School of Medicine and analyzed using QIIME 2 (v2024.5) software, with over 10,000 paired-end 150 ×150 base pair (bp) reads. 16S-rRNA libraries were prepared using the Earth Microbiome Project protocol with primers as described by Walters et al.[1] Specific bacterial communities, sequences were classified into operational taxonomic units (OTUs) and categorized by individual microbial taxa, including phylum, class, order, family, and genus clusters. Counts of less than 100 OTUs were excluded. Bar-graph plots were created based on the relative abundance of OTUs. To further study similarity (or dissimilarity) in terms of microbiota composition among samples, a principal coordinate ordination based on weighted UniFrac distances was generated. Specific bacterial communities' sequences were classified into operational taxonomic units and categorized by individual microbial taxa, including phylum, class, order, family, and genus clusters.

**Sequence Processing and Taxonomic Classification**

To remove host-derived sequences, raw paired-end reads were taxonomically classified using Kraken2 (v2.1.2)[2], a k-mer–based taxonomic classifier. A custom Kraken2 database was built using the Mus musculus reference genome (GRCm39, RefSeq assembly accession GCF_000001635.27) and associated taxonomy files from the NCBI RefSeq repository. For each sample, sequences classified as mouse were separated and discarded, while unclassified reads—representing putative microbial sequences—were retained for microbiome analysis.

Microbial sequence quality control and amplicon sequence variant (ASV) inference were performed using the DADA2 plugin within QIIME 2 (v2024.5)[3,4]. Although paired-end sequencing was conducted, the reverse reads exhibited consistently low quality scores (Phred < 30), especially towards the 3′ ends. After trimming, the overlap between forward and reverse reads was insufficient for reliable merging (successful merging rate<10%). Therefore, only forward reads were retained and processed as single-end sequences. Denoising was conducted using DADA2 with the first 10 bases trimmed and reads truncated at 180 bp to remove low-quality tails. Chimeric sequences were identified and removed using the consensus method.

ASVs were clustered into OTUs at 99% identity against the SILVA 138 reference database[5] using the VSEARCH algorithm implemented in QIIME 2 (via the vsearch cluster-features-closed-reference method)[6]. Unmatched sequences were excluded from downstream analysis. Taxonomic classification was performed on representative sequences using a Naive Bayes classifier pre-trained on the SILVA 138 99% reference via QIIME 2's feature-classifier classify-sklearn plugin[7].

**Statistical Analysis**

All diversity analyses were conducted in R (v4.4.1). Alpha diversity was assessed using the Shannon diversity index and Chao1 richness estimator. The Shannon index was computed using the diversity() function and Chao1 values were obtained via the estimateR() function from the vegan package (v2.6-4)[8]. For group-wise comparisons, Wilcoxon rank-sum tests were used to evaluate differences in alpha diversity between brain and cecum samples, as well as between acute and chronic samples. Beta diversity was evaluated using the weighted UniFrac distance.[9,10] Principal coordinates analysis (PCoA) was used for visualization, performed using the cmdscale() function in R. Comparisons between sample groups (brain vs. cecum; acute vs. chronic) were tested using PERMANOVA implemented via the adonis() function from the vegan package.

Differential abundance analysis was conducted using the microbiomeMarker R package (v1.2.0)[11] implementing the LEfSe (Linear discriminant analysis Effect Size) algorithm[12]. Microbial features enriched between brain and cecum samples were identified using a default LDA score cutoff of 2.0 and a significance threshold of 0.1 (after FDR adjustment). To further explore group-specific microbial enrichment patterns, LEfSe analyses were independently performed within brain and cecum subsets. Within each subset, microbial profiles of negative control samples were used as background to identify taxa enriched in the trauma control, PSC implantation and AIC implantation subgroups, respectively. Because each subgroup contained only 6 mice, meaningful statistical significance of differential abundance analysis cannot be expected. Rather than treating p-values as evidence, we used a lenient filter ($p < 0.2$) and emphasized LDA scores to provide a descriptive ranking of taxa showing the largest shifts. These subgroup findings should be viewed as exploratory patterns intended to aid biological interpretation, rather than as formal statistical evidence of differential abundance.

Functional profiles of the microbial communities were predicted from 16S rRNA gene sequences using PICRUSt2.[13] Amplicon sequence variants (ASVs) and their abundances were provided as inputs in FASTA and BIOM format, respectively. The default PICRUSt2 pipeline (picrust2_pipeline.py) was run with 12 threads and the --stratified option to obtain per-taxon contributions to predicted functional pathways. KEGG Orthologs (KOs) were used as the primary functional units for downstream analysis. Specific KO identifiers related to short-chain fatty acid (SCFA) biosynthesis and lipopolysaccharide (LPS) biosynthesis pathways were selected for focused analysis across experimental groups. Within both the brain and cecum subsets, we compared the predicted abundances of these functional pathways among unaltered control, trauma control, PSC implantation, and AIC implantation groups using boxplots. Effect size is calculated using Cliff's Delta and a permutation p-value is provided.[14] To further explore the microbial contributors to functional shifts, bubble plots were generated to visualize the associations between taxa with high LDA scores (as identified by LEfSe) and selected functional pathways within each subgroup. This integrative approach allowed us to link taxonomic enrichment to potential functional consequences in the context of implant-associated inflammation.

**In vivo MR imaging**

*In-vivo* MRI was performed on five mice: negative control (n = 1); mice implanted with PSC (n = 2); and mice implanted with AIC (n = 2). MRI was performed on each mouse at one, four, eight-, and 16-weeks post-implantation.

All MRI scans were performed on a horizontal bore 11.7 Tesla (11.7T) Bruker Biospec MRI system (Bruker, Ettlingen, Germany) equipped with a physiological monitoring system. A 72-mm quadrature volume resonator was used as a transmitter, and a 4-element (2 × 2) phased array coil was used as a receiver. Anesthesia was induced with 2% isoflurane in medical air, followed by 1% isoflurane in oxygen and air (1:3 ratio) for maintenance during the MRI scan, applied via a vaporizer and a facial mask. During the MRI scans, the mouse head was positioned using a bite bar and 2 ear pins, and the animal was placed on a water-heated bed equipped with temperature and respiratory controls. Respiration rate and body temperature were monitored constantly by an animal monitoring system (SAII, Stony Brook, NY, USA) during MRI. Animals were ventilated to maintain stable physiological conditions (respiratory rate 50–55 breaths/min).

The following MRI scans were performed on each animal during each session: 1) Anatomical images were acquired using a T2-weighted multi-slice rapid acquisition with relaxation enhancement (RARE) sequence: fast spin echo readout, echo time (TE)/repetition time (TR) = 30/2500 ms, resolution = $0.06 \times 0.06 \times 1.0$ mm3, 15 slices, RARE factor = 8; 2) Fluid attenuated inversion recovery (FLAIR) MRI: fast spin echo readout, inversion time (TI) = 1427 ms, echo time (TE)/repetition time (TR) = 30/3000 ms, resolution = $0.06 \times 0.06 \times 1.0$ mm3, 15 slices, RARE factor = 8; 3) Multi-echo susceptibility weighted imaging (SWI): fast gradient echo readout, echo time (TE)/repetition time (TR) = 4,10,16,23,29,35/2,000 ms, resolution = $0.03 \times 0.03 \times 0.8$ mm3, 19 slices. For this MRI scan, each animal received an intravenous (IV) injection (via the tail vein) of an ultra-small-superparamagnetic-iron-oxides (USPIO) contrast agent: Ferumoxytol (30mg/mL, Feraheme, Waltham, MA, USA), the dose was calculated based on the standard dosage used in human studies and the animals' body weight. The SWI MRI scans were performed approximately 24 hours after the contrast injection. Animals were then euthanized.

**MRI data analysis**

Image analysis was performed using the statistical parametric mapping software package (Version 8, Welcome Trust Centre for Neuroimaging, London, UK; http://www.fil.ion.ucl.ac.uk/spm/) and other in-house codes written in Matlab (MathWorks, Natick, MA, USA). All MR images from each mouse were co-registered. Four regions of interest (ROI) were manually delineated in each mouse based on T2-weighted RARE and FLAIR images using ITK-SNAP software (**Fig. xx**). First, the implant ROI was identified as the region with signal intensities below one standard deviation of background signals in both T2-weighted RARE and FLAIR images in the brain. Second, the edema ROI was traced around the implant ROI. As fluid signals show high intensities in T2-weighted RARE images but are suppressed in FLAIR scans, edema should show signal intensities greater than average parenchymal signals in T2-weighted RARE images and signal intensities comparable to average parenchymal signals in FLAIR images. Thirdly, glial scar ROI was delineated around the implant ROI. It should show comparable signal intensities in both T2-weighted RARE and FLAIR images that are slightly lower than average parenchymal signals. Finally, a ROI of normal appearing tissue on the contralateral side of the implant was manually drawn. Volumes of each ROI were estimated from anatomical images. (**Fig. 2**) $R2^*$ maps were calculated from the SWI images. Average $R2^*$ values were calculated in each ROI. $R2^*$ values are an estimation of macrophage activity, which should be present in the glial scar region but should not change in the implant region or contralateral brain.